\pdfoutput=1
\documentclass{PoS}

\title{Precise neutron lifetime experiment using pulsed neutron beams at J-PARC}

\ShortTitle{Precise neutron lifetime experiment using pulsed neutron beams at J-PARC}

\author{\speaker{Naoki Nagakura}$^{1}$, Katsuya Hirota$^{2}$, Sei Ieki$^{1}$, Takashi Ino$^{3}$, Yoshihisa Iwashita$^{4}$, Masaaki Kitaguchi$^{5}$, Ryunosuke Kitahara$^{6}$, Kenji Mishima$^{3}$, Aya Morishita$^{7}$, Hideyuki Oide$^{8}$, Hidetoshi Otono$^{9}$, Risa Sakakibara$^{2}$, Yoshichika Seki$^{10}$, \ \ \ \ \ \ \ \  Tatsushi Shima$^{11}$, Hirohiko M. Shimizu$^{2}$, Tomoaki Sugino$^{2}$, Naoyuki Sumi$^{7}$, Hirochika Sumino$^{12}$, Kaoru Taketani$^{3}$, Genki Tanaka$^{7}$, Tatsuhiko Tomita$^{7}$,  \ \ \ \ \ \ \ Takahito Yamada$^{1}$, Satoru Yamashita$^{13}$, Mami Yokohashi$^{4}$, and Tamaki Yoshioka$^{9}$\\
      $^{1}$Department of Physics, Graduate School of Science, The University of Tokyo \\
      $^{2}$Department of Physics, Graduate School of Science, Nagoya University\\
	$^{3}$KEK (High Energy Accelerator Research Organization) \\
	$^{4}$Institute for Chemical Research, Kyoto University\\
	$^{5}$Kobayashi-Maskawa Institute for the Origin of Particles and the Universe (KMI), Nagoya University\\
	$^{6}$Department of Physics, Graduate School of Science, Kyoto University\\
	$^{7}$Department of Physics, Graduate School of Science, Kyushu University\\
	$^{8}$Istituto Nazionale di Fisica Nucleare (INFN-Genova)\\
	$^{9}$Research Center for Advanced Particle Physics (RCAPP), Kyushu University\\
	$^{10}$Neutron Science Section, Materials and Life Science Division, J-PARC Center, Japan Atomic Energy Agency\\
	$^{11}$Research Center for Nuclear Physics (RCNP), Osaka University\\
	$^{12}$Department of Basic Science, Graduate School of Arts and Sciences, The University of Tokyo\\
	$^{13}$International Center for the Elementary Particle Physics (ICEPP), The University of Tokyo\\
       E-mail: \email{nagakura@icepp.s.u-tokyo.ac.jp}}

\abstract{
The neutron lifetime ($\tau_{n}$) is one of the basic parameters in the weak interaction, and is used for predicting the light element abundance in the early universe. Our group developed a new setup to measure $\tau_{n}$ with the goal precision of 0.1\% at the polarized beam branch BL05 of MLF, J-PARC. The commissioning data was acquired in 2014 and 2015, and the first set of data to evaluate $\tau_{n}$ in 2016, which is expected to yield a statistical uncertainty of $\mathcal{O}(1)\%$. This paper presents the current analysis results and the future plans to achieve our goal precision.
}

\FullConference{The 26th International Nuclear Physics Conference\\
		11-16 September, 2016\\
		Adelaide, Australia}

\begin{document}

\section{Introduction}
A free neutron decays into a proton, an anti-neutrino, and an electron through the beta decay process, and its lifetime ($\tau_{n}\sim\ $880 sec) is a fundamental parameter in the weak interaction. In the theory of the Big Bang Nucleosynthesis, $\tau_{n}$ plays a crucial role in predicting the light element abundance in the early universe \cite{cite:bbn}. It is also used for evaluating the $V_{\mathrm{ud}}$ element in the CKM matrix \cite{cite:pdg}. Our group developed a new setup to measure $\tau_{n}$ with a precision comparable to that of previous experiments, using the pulsed neutron beam at J-PARC (Japan Proton Accelerator Research Complex).
\subsection{Motivation}
\begin{figure}
 \centering
 \includegraphics[width=0.6\columnwidth]{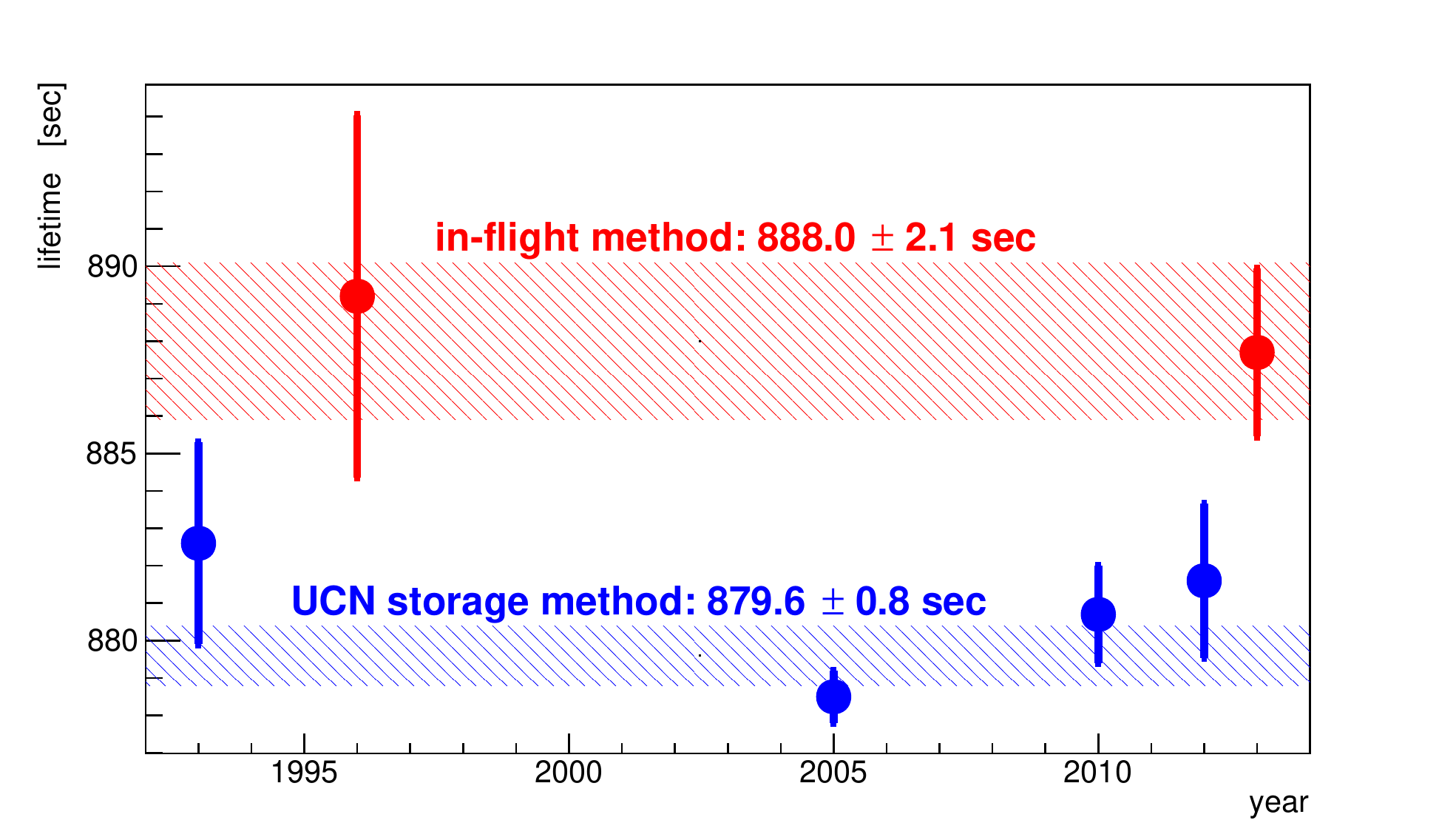}
 \caption{Results of the previous neutron lifetime experiments \cite{cite:pdg}\cite{cite:lifetime}. The results of the in-flight method and the UCN storage method are expressed as red and blue points, respectively. There exists a 1.0\% ($3.8\sigma$) deviation between the average results of these two methods.}
 \label{fig:lifetime}
\end{figure}

There are mainly two types of previous experiments to measure $\tau_{n}$. One is the in-flight method, in which a cold neutron beam is injected into a chamber and the protons from beta decay are detected. The other is the UCN storage method, in which UCNs (Ultra Cold Neutrons) are stored in a chamber and the number of remaining neutrons is counted after a certain period of time. However, there exists a significant deviation of 1.0\% (3.8$\sigma$) between the results of these two methods (Figure \ref{fig:lifetime}). This prevents, for example, the precise prediction of the light element synthesis in the early universe. In order to test which result is correct, our group plans to measure $\tau_{n}$ with the precision of 0.1\% using a different method. 

\subsection{Measurement principle}
As the detector of this experiment, a Time Projection Chamber (TPC) filled with He and CO$_2$ gas is used. The pulsed neutron beam enters the TPC and both electrons from beta decay and $^3$He(n, p)$^3$H events are detected simultaneously. Since the $^3$He pressure is evaluated in advance, the total neutron fluence can be evaluated using the number of $^3$He(n, p)$^3$H events detected by the TPC. Finally, $\tau_{n}$ can be expressed as the ratio of these two types of events as
\begin{equation}
	\tau_{n} = \frac{1}{\sigma (v_{0})v_{0}\rho}\frac{(N_{^{3}\mathrm{He}}/\varepsilon_{^{3}\mathrm{He}})}{(N_{\beta}/\varepsilon_{\beta})} .
\end{equation}
Here, $\sigma(v_{0})$ is the $^3$He(n, p)$^3$H cross section at the neutron velocity of $v_{0}=2200$ m/s, $\rho$ is the number density of $^3$He in the TPC. We utilize the characteristic that the cross section is inversely proportional to the neutron velocity. $N_{^{3}\mathrm
{He}}$ and  $N_{\beta}$ represent the number of detected events in the TPC of $^3$He(n, p)$^3$H and beta decay, respectively, and $\varepsilon_{^{3}\mathrm{He}}$ and $\varepsilon_{\beta}$ are their respective efficiencies. Each efficiency is evaluated using the Monte Carlo simulation based on the Geant4 simulation package \cite{cite:geant4}.

\section{Setup}
This experiment utilizes a pulsed neutron beam at BL05, MLF (Materials and Life Science Experimental Facility), J-PARC. Here, the neutron beam has a typical energy of 10 meV, and it is polarized to the ratio of 95.62(3)\% \cite{cite:ino}. The expected neutron flux at this polarized beam branch is $(3.9\pm 0.3)\times 10^{7}$ $/\mathrm{cm}^{2}/\mathrm{s}$ at the design operation power of 1 MW \cite{cite:beamline1}\cite{cite:beamline2}.
The polarized neutron beam first enters a device called the Spin Flip Chopper to form neutron bunches. Downstream of the Spin Flip Chopper is the TPC, which detects both beta decay and $^3$He(n, p)$^3$H events simultaneously. The length of each bunch is made to be about half of the TPC length, therefore we are able to analyze the data only when the neutron bunch is completely contained inside the TPC.

\subsection{Spin Flip Chopper}
\begin{figure}
 \centering
 \includegraphics[width=0.6\columnwidth]{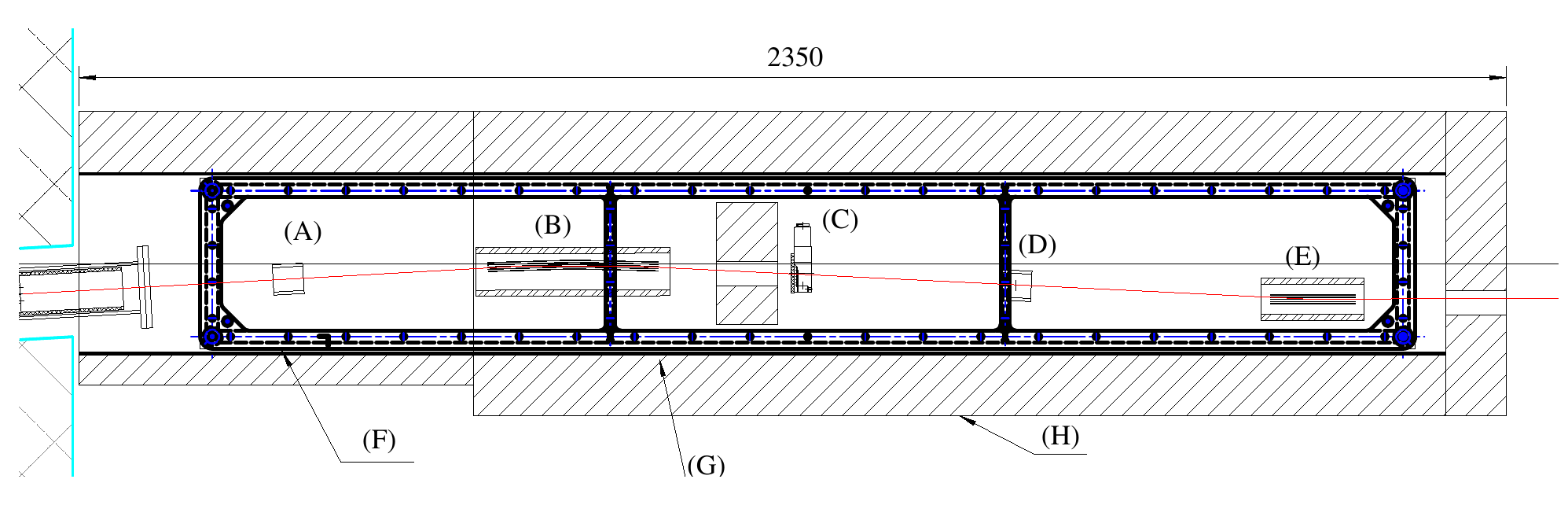}
 \caption{Schematic view of the Spin Flip Chopper. (A)(D): RF flippers, (B)(E): magnetic super mirrors. These devices are covered by Pb shield.}
 \label{fig:sfc}
\end{figure}

The Spin Flip Chopper is used for forming neutron bunches from the continuous neutron beam \cite{cite:sfc}. It is composed of two RF flippers and three magnetic super mirrors (Figure \ref{fig:sfc}). A magnetic field of 1 mT is applied over all devices to maintain the neutron polarization. The RF flipper is able to flip the neutron polarization when the flipping frequency satisfies the resonance condition. In the magnetic super mirror, the potential a neutron feels changes depending on its polarization direction, thus only neutrons polarized in a particular direction are reflected and guided to the downstream TPC. Therefore, by adjusting the duty factor of these flippers, the neutron bunches with arbitrary length can be formed.
\subsection{Time Projection Chamber}
\begin{figure}
 \centering
 \includegraphics[bb=0 0 840 1100,width=0.55\columnwidth]{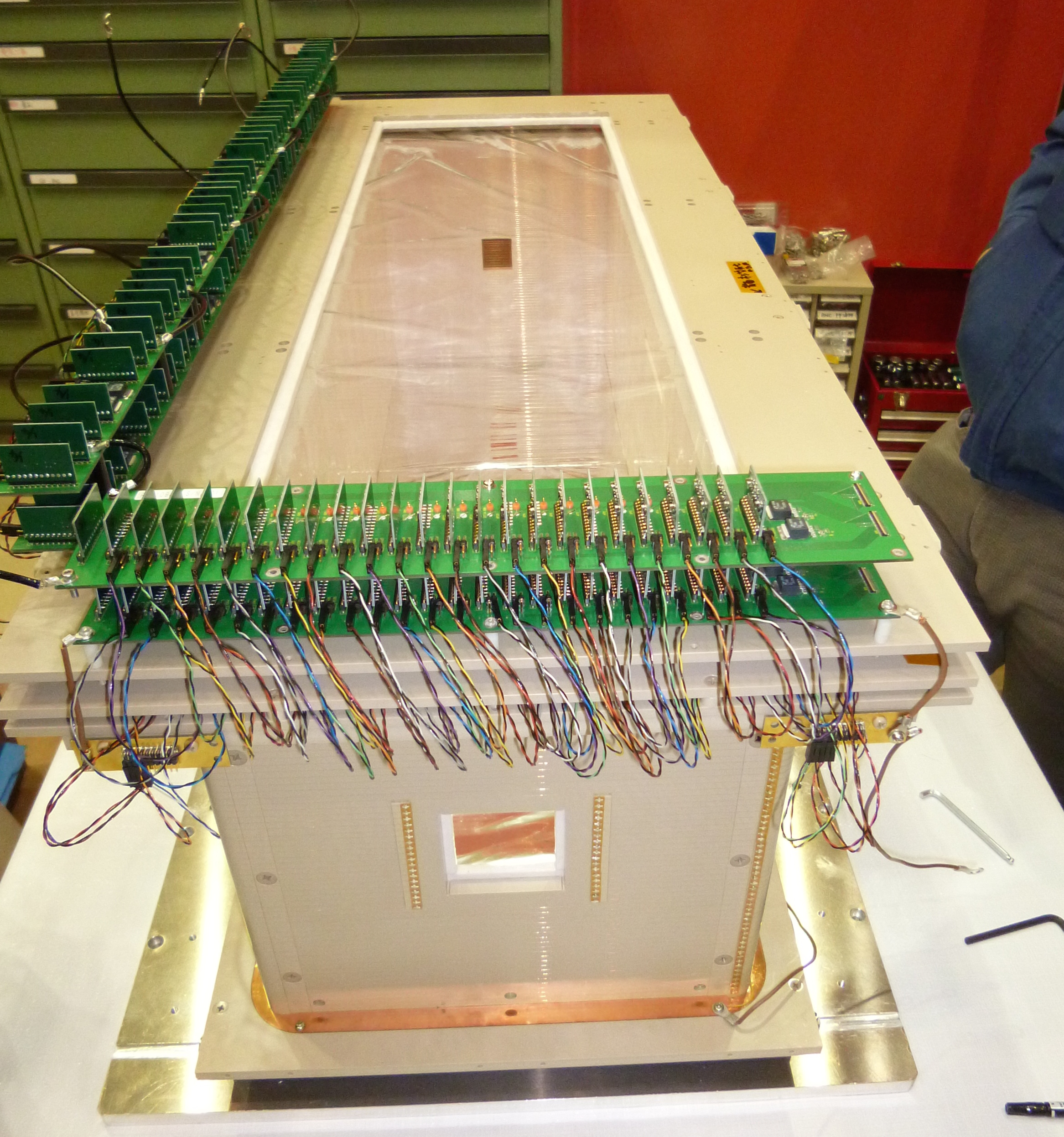}
 \caption{Photo of the Time Projection Chamber.}
 \label{fig:tpc}
\end{figure}

The TPC has the sensitive region of 290 mm $\times$ 300 mm $\times$ 960 mm (Figure \ref{fig:tpc}), filled with $^4$He (85 kPa) + $^3$He (100 mPa) + CO$_2$ (15 kPa) gas \cite{cite:tpc}. There are three layers of MWPC (Multi Wire Proportional Chamber) in the upper part to collect drift electrons. Anode and field wires are arranged alternately in the same plane parallel to the beam direction. Cathode wires are strung above and below the layer of anode and field wires, perpendicular to the beam direction. The wire interval is 6 mm in all planes. Signals at each wire are amplified in preamplifiers and read out through 12-bit flash ADCs. 
The electric field inside the TPC is set about 300 V/cm, and the corresponding drift velocity of electrons is evaluated to be 1.0 $\mathrm{cm/\mu s}$.\par
The TPC has two important characteristics specialized for this experiment. One is that it is made of low background material called PEEK (PolyEthel Ethel Ketone). This significantly reduces the environmental background for beta decay. The other is that its inside walls are covered by 5 mm-thick $^6$LiF plates in order to absorb scattered neutrons without the emission of prompt gamma rays.
\section{Analysis}
This section describes the analysis strategies to evaluate $\tau_{n}$ from the acquired data. First, the whole data set for the analysis is introduced. Next, the separation of the two types of signal events is explained. Finally, subtraction methods are proposed for each type of backgrounds.
\subsection{Data set}
\begin{figure}
 \centering
 \includegraphics[width=1\columnwidth]{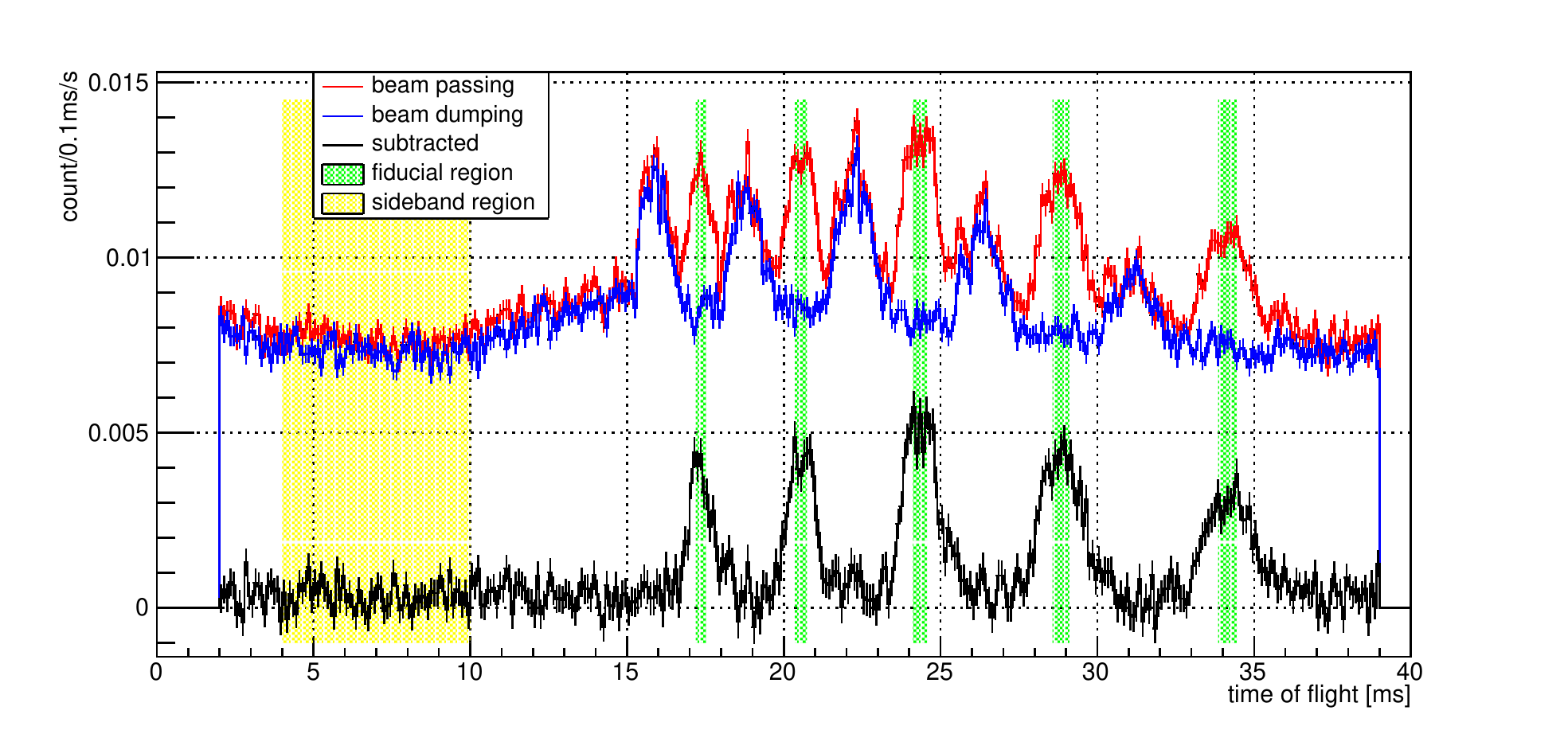}
 \caption{Time of flight distribution for the beam passing and the beam dumping modes. The subtracted distribution is also shown.}
 \label{fig:sn}
\end{figure}

Table \ref{table:datacycle} lists the data set acquired during the beam time in 2016. In addition to the beam passing mode, the data of the beam dumping mode is acquired for evaluating background, during which a $^{6}$LiF shutter immediately upstream of the TPC is closed. X-rays of 5.9 keV from a $^{55}$Fe source are used for the energy calibration. Cosmic ray data is used for evaluating the drift velocity of electrons. The time of flight distribution for both the beam passing and the beam dumping modes are shown in Figure \ref{fig:sn}.\\

\begin{figure}
 \centering
 \includegraphics[width=0.8\columnwidth]{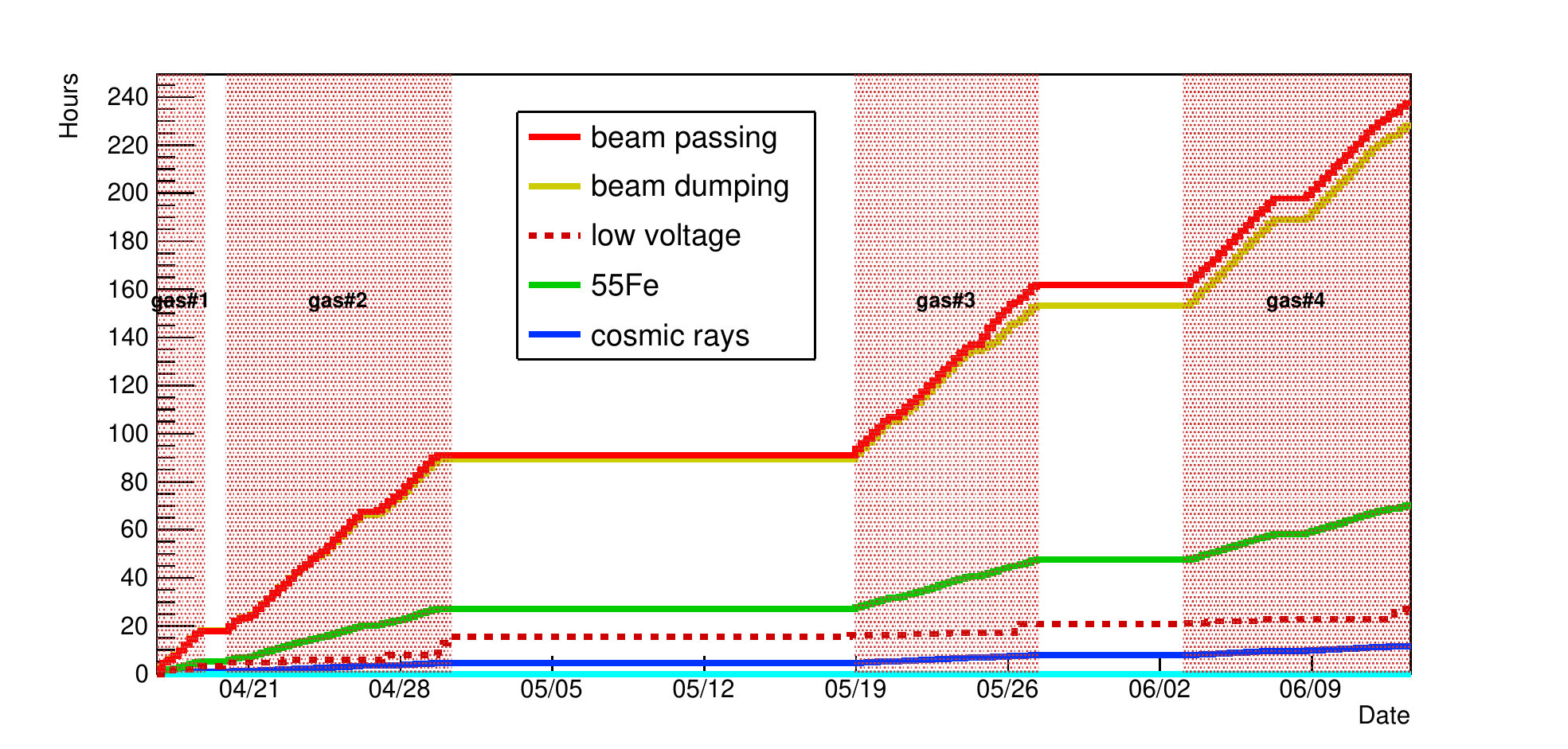}
 \caption{Result of data acquisition in 2016. The combined data taking time reached 250 hours for both the beam passing and the beam dumping modes.}
 \label{fig:status}
\end{figure}

\begin{table}
 \caption{Data cycle}
 \label{table:datacycle}
 \centering
 \begin{tabular}{|l|c|c|l|}
  \hline
 mode & shutter status & time [sec] & purpose \\ \hline
 beam passing & open & 1000 & count the numbers of beta decay and $^3$He(n, p)$^3$H events \\
 beam dumping & close & 1000 & for background measurement  \\
 $^{55}$Fe X-rays & close & 300 & energy calibration  \\
 cosmic rays & close & 100 & drift velocity calibration  \\ \hline
 low gain & open & 1000 & evaluate the outgass pressure \\ \hline
 \end{tabular}
\end{table}
The commissioning data was acquired in 2014 and 2015, and the first set of data to evaluate $\tau_{n}$ from April to June in 2016. Figure \ref{fig:status} shows the result of data accumulation in 2016 with four types of gas mixture. The total data taking time reached about 250 hours for both the beam passing and beam dumping modes. The MLF operation power was about 200 kW at this time. The combined statistical uncertainty on $\tau_{n}$ is expected to be about 1\%.\\
\subsection{Separation of $^3$He(n, p)$^3$H and beta decay events}
The precise separation of $^3$He(n, p)$^3$H and beta decay events are significant in this experiment because it directly affects the result of $\tau_{n}$. Since the energy deposit per wire for $^3$He(n, p)$^3$H is much higher than that of beta decay, these two can be separated using the maximum energy deposit among all field wires. \par
\begin{figure}
 \centering
 \includegraphics[width=1\columnwidth]{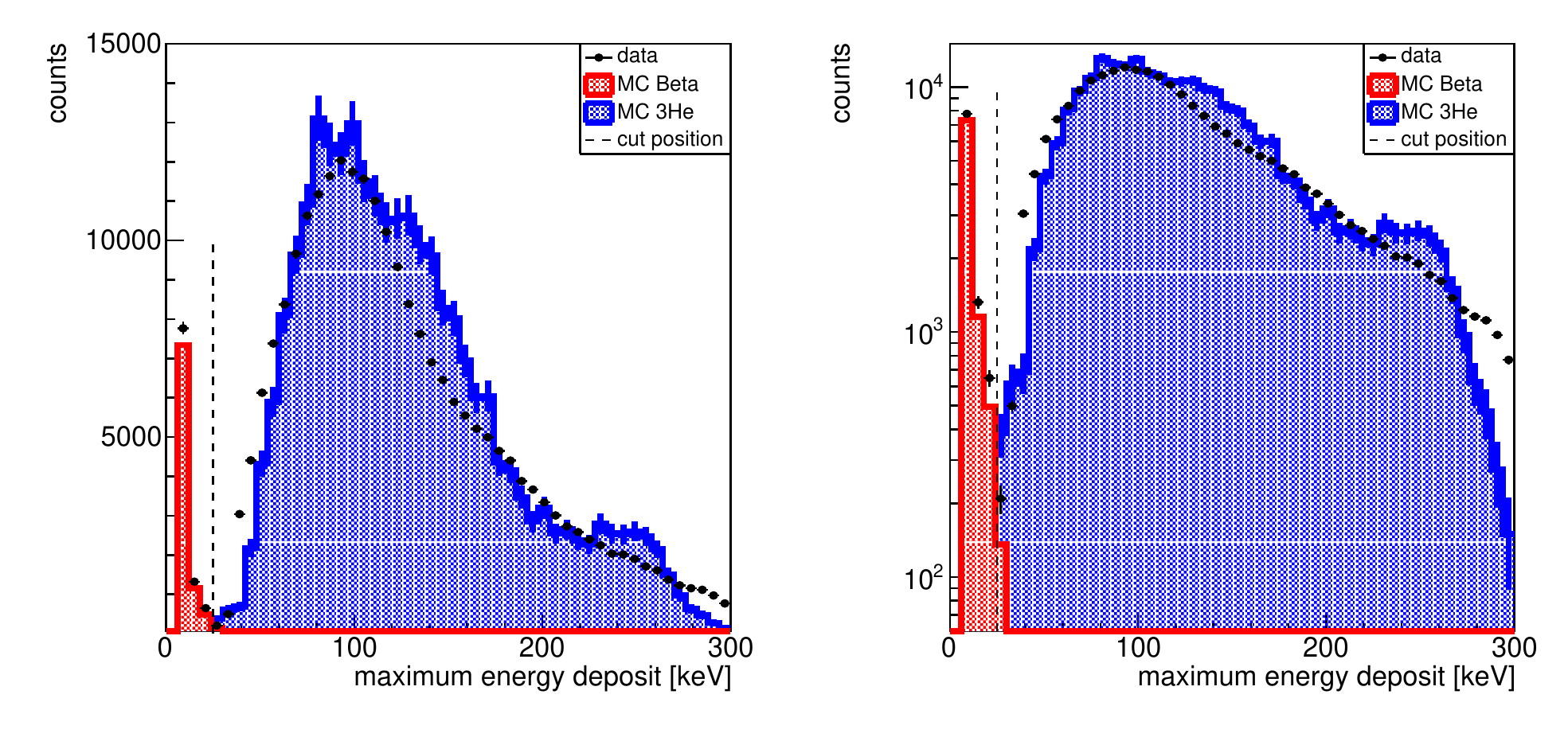}
 \caption{Distribution of the maximum energy deposit among all field wires; linear scale (left) and semilogarithmic scale (right). The separation energy between beta decay and $^3$He(n, p)$^3$H is set at 25 keV.}
 \label{fig:fphmax}
\end{figure}

Figure \ref{fig:fphmax} shows the distribution of the maximum energy deposit among all field wires for both the beta decay and $^3$He(n, p)$^3$H events. The separation energy is set at 25 keV, at which the selection efficiencies of 97.98(5)\% and 99.98\% can be obtained for beta decay and $^3$He(n, p)$^3$H, respectively.
\subsection{Background}
There are four types of backgrounds expected for beta decay. One is the environmental background (A), such as cosmic rays and natural radiation. Another is referred to as the upstream background (B), which is the radiation originating mainly from the neutron source and the Spin Flip Chopper. The third is the radioactivation inside the TPC (C). The fourth is the beam-induced background (D), such as prompt gamma rays coming from the interaction between the TPC wall and a scattered neutron. \par
\begin{figure}
 \centering
 \includegraphics[width=1\columnwidth]{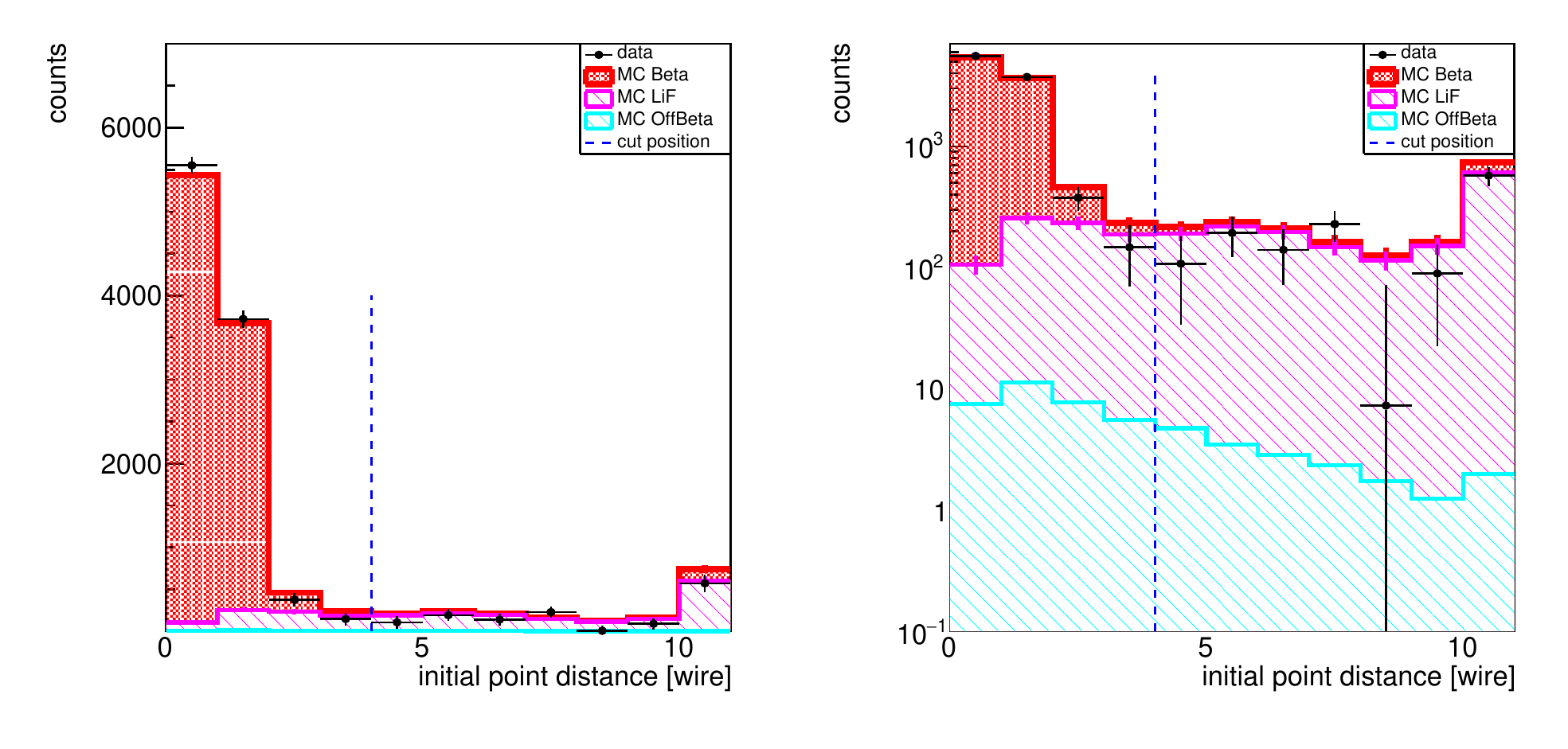}
 \caption{Distribution of the distance between the closer endpoint of the track and the beam position; linear scale (left) and semilogarithmic scale (right). The cut position is set at the distance of 4 wires.}
 \label{fig:anddc}
\end{figure}

Both type(A) and type(B) backgrounds can be subtracted using the beam dumping data because they exist even without neutrons inside the TPC. Type(C) background can be estimated using the sideband region, where the time of flight of a neutron is between 4 ms and 10 ms. Type(D) background can be evaluated based on the characteristic that tracks exist uniformly inside the TPC. Figure \ref{fig:anddc} shows the distribution of the distance between the closer endpoint of the track and the beam position. The cut position is set at the distance of 4 wires (24 mm) from the beam center. The efficiency for beta decay at this cut position is evaluated to be 97.56(5)\%. The event rate in the background region is interpolated to estimate the amount of type(D) background inside the signal region.
\subsection{List of uncertainties}
Table \ref{table:uncertainties} lists the uncertainties estimated for $\tau_{n}$ using the commissioning data acquired in 2014. All types of uncertainties have been evaluated with the exception of the beam-induced background, which is undergoing study. After the completion of our analysis, the first result will be published soon, whose precision is expected to be $\mathcal{O}(1)\%$.
\begin{table}
 \caption{Current list of corrections and uncertainties for $\tau_{n}$.}
 \label{table:uncertainties}
 \centering
 \begin{tabular}{|l|l|rr|}
  \hline
  parameter &  & uncertainty [\%] & correction [\%] \\ \hline
 $N_{\beta}$ & statistics & $\sim$1 & ---\\ 
   & $^3$He(n, p)$^3$H leakage & $<$\ 0.34& 0 \\
   & beam-induced background & being evaluated & 8.6 \\
   & efficiency & $^{+1.0}_{-0.3}$ &6.1 \\
   & pileup & 0.39& -0.39 \\
   & background subtraction & 0.28 & -0.43 \\ \hline
 $N_{^{3}\mathrm{He}}$ & $^{14}$N background & 0.23 & -1.45 \\
   & $^{17}$O background & 0.03 & -0.50 \\ \hline
 $N_{\beta}$ and $N_{^{3}\mathrm{He}}$ & Spin Flip Chopper S/N & $<$ 0.5 & $<$ 0.5 \\ \hline
 $\rho$ & pressure measurement& 0.65 & ---\\
   & chamber deformation (pressure) & $<$\ 0.33 & -0.33 \\
   & temperature non-uniformity & 0.23 & 0.23 \\ \hline
 $\sigma$ & $^3$He(n, p)$^3$H cross section & 0.13 & --- \\ \hline
 \end{tabular}
\end{table}
\section{Future plan}
In order to achieve our goal precision of 0.1\%, both statistical and systematic uncertainties need to be reduced by a factor of 10. Regarding statistics, the MLF operation power will increase from current 200 kW to the design operation power of 1 MW within a few years. In addition, we plan to upgrade the beam transportation system so that the whole beam at this beam branch (4 cm × 10 cm) can be injected into the detector. We expect to achieve our goal precision from 40 days data taking after these upgrades. Regarding systematics, it is also planned to add additional layers of wires in the TPC in order to improve the separation of the beam-induced background.
\section{Conclusion}
The neutron lifetime is one of the fundamental parameters in the weak interaction. However there exists a significant deviation of $3.8\sigma$ between the results of previous two types of experiments. In order to resolve this serious problem, our group developed the new devices to measure $\tau_{n}$ using the pulsed neutron beam at J-PARC. Almost all of the uncertainties have been evaluated at present, and the first result, whose precision is $\mathcal{O}(1)\%$ on $\tau_{n}$, will be published soon. Several upgrade plans to achieve our goal precision of 0.1\% are also being carried out in our group.

\end{document}